\documentclass[prd,showpacs,showkeys,12pt,a4paper]{revtex4-1}
\usepackage{cancel,amsmath,amssymb,
graphicx}
\usepackage{amsfonts,amsthm
}
\usepackage{latexsym}

\def\stamp{--- {\bf \today} --- {\bf \jobname.tex}}

\def\BE{\begin{equation}}
\def\EE{\end{equation}}
\def\spa#1.#2{\left\langle#1\,#2\right\rangle}
\def\spb#1.#2{\left[#1\,#2\right]}
\def\spba#1.#2.#3{\left[#1|#2|#3\right\rangle}
\def\spab#1.#2.#3{\left\langle#1|#2|#3\right]}
\def\spaa#1.#2.#3{\left\langle#1|#2|#3\right\rangle}
\def\spbb#1.#2.#3{\left[#1|#2|#3\right]}
\def\lor#1.#2{\left(#1\,#2\right)}

\def\Year{\expandafter\eatPrefix\the\year}
\newcount\hours \newcount\minutes
\def\monthname{\ifcase\month\or
January\or February\or March\or April\or May\or June\or July\or
August\or September\or October\or November\or December\fi}
\def\shortmonthname{\ifcase\month\orx
Jan\or Feb\or Mar\or Apr\or May\or Jun\or Jul\or
Aug\or Sep\or Oct\or Nov\or Dec\fi}

\def\TimeStamp{\hours\the\time\divide\hours by60%
\minutes -\the\time\divide\minutes by60\multiply\minutes by60%
\advance\minutes by\the\time%
${\rm \shortmonthname}\cdot   \if\day<100\fi\the\day\cdot   \the\year
\qquad\the\hours:\if\minutes<100\fi\the\minutes$}

\newskip\humongous \humongous=0pt plus 1000pt minus 100pt

\newif\ifdtup

\newcounter{eqnumber}[section]

\newbox\charbox
\newbox\slabox

\def\spa#1.#2{\left\langle#1\,#2\right\rangle}
\def\spb#1.#2{\left[#1\,#2\right]}
\def\lor#1.#2{\left(#1\,#2\right)}

\catcode`@=11  

\def\half{{1\over 2}}

\def\lsl{\not{\hbox{\kern-2.3pt $\ell$}}}
\def\ksl{\not{\hbox{\kern-2.3pt $k$}}}

\def\spa#1.#2{\left\langle#1\,#2\right\rangle}
\def\spb#1.#2{\left[#1\,#2\right]}
\def\lor#1.#2{\left(#1\,#2\right)}

\def\sand#1.#2.#3{%
  \left\langle\smash{#1}{\vphantom1}\right|{#2}%
  \left|\smash{#3}{\vphantom1}\right\rangle}
\def\sandp#1.#2.#3{%
  \left\langle\smash{#1}{\vphantom1}^{-}\right|{#2}%
  \left|\smash{#3}{\vphantom1}^{+}\right\rangle}
\def\sandpp#1.#2.#3{%
  \left\langle\smash{#1}{\vphantom1}^{+}\right|{#2}%
  \left|\smash{#3}{\vphantom1}^{+}\right\rangle}
\def\sandmm#1.#2.#3{%
  \left\langle\smash{#1}{\vphantom1}^{-}\right|{#2}%
  \left|\smash{#3}{\vphantom1}^{-}\right\rangle}
\def\sandpm#1.#2.#3{%
  \left\langle\smash{#1}{\vphantom1}^{+}\right|{#2}%
  \left|\smash{#3}{\vphantom1}^{-}\right\rangle}
\def\sandmp#1.#2.#3{%
  \left\langle\smash{#1}{\vphantom1}^{-}\right|{#2}%
  \left|\smash{#3}{\vphantom1}^{+}\right\rangle}

\begin{document}

\title{Gravitino Interactions from Yang-Mills Theory\bigskip\bigskip\bigskip}

\author{N.~E.~J.~Bjerrum-Bohr and Oluf~Tang~Engelund}\bigskip
\affiliation{
Niels Bohr International Academy,\\
The Niels Bohr Institute,\\
Blegdamsvej 17, DK-2100 Copenhagen \O,
Denmark\bigskip\bigskip\bigskip\bigskip}\bigskip
\email{bjbohr@nbi.dk; engelund@nbi.dk}

\begin{abstract}
We fabricate gravitino vertex interactions using as only input
on-shell Yang-Mills amplitudes and the Kawai-Lewellen-Tye gauge
theory / gravity relations. A useful result of this analysis is
simpler tree-level Feynman rules for gravitino scattering than
in traditional gauges. All results are explicitly verified until
five point scattering. \\ \\
 \end{abstract}\bigskip

\date{\today}

\maketitle

\section{Introduction}
A remarkable and intimate connection between gravity
and Yang-Mills amplitudes exist through the Kawai-Lewellen-Tye
(KLT) relations~\cite{Kawai:1985xq}. In the field theory limit
($\alpha'\rightarrow 0$) these relations take the form
\begin{equation}
{\cal M_{\rm Gravity}} \sim \sum_{ij} K_{ij} {{\cal A}^{i\; L}_{\rm Yang-Mills}}
\times { {{\cal A}^{j\; R}_{\rm Yang-Mills}}}\,.
\end{equation}
Here $\mathcal{M}_{\rm Gravity}$, ${\cal A}^{i\; L}_{\rm
Yang-Mills}$, ${\cal A}^{j\;R}_{\rm Yang-Mills}$ are gravity
and Yang-Mills amplitudes and $K_{ij}$ is a specific function
of kinematic invariants depending on the input Yang-Mills
amplitudes. The KLT relations are a widely used tool for
computing gravity amplitudes via a recycling of results for
Yang-Mills amplitudes. For tree and loop amplitudes KLT can be
combined with string theory inspired diagram rules as well as
unitarity cut techniques~\cite{Bern:1993wt,BDDPR,Bern:1999ji}.
Recently, inspired by Witten's
work~\cite{WittenTopologicalString}, we have seen remarkable
progress in Yang-Mills amplitude computations (for recent
reviews see~\cite{Cachazo:2005ga,Bern:2007dw}). Via KLT, this
progress has fed into gravity yielding a much better
understanding of
amplitudes~\cite{Giombi:2004ix,Bern:2005bb,BBSTgravity,BDIgravity,
BjerrumBohr:2005jr,BjerrumBohr:2006yw,Benincasa:2007qj,Bern:2007xj,Ananth:2007zy,Elvang:2007sg,Hall:2008xn,
BjerrumBohr:2008vc,ArkaniHamed:2008yf,BjerrumBohr:2008ji,Bern:2008qj,ArkaniHamed,Badger:2008rn,Katsaroumpas:2009iy,
Bern:2006kd,Cheung:2009dc,BjerrumBohr:2009rd,Stieberger:2009hq,Boels:2009bv,Chen:2010sr}.
It is now clear that gravity tree amplitudes have additional
simplicity in their expressions. In many cases this simplicity
appears very naturally from a string theory
viewpoint~\cite{BjerrumBohr:2008vc,
BjerrumBohr:2008ji,ArkaniHamed,
Bern:2008qj,Badger:2008rn,BjerrumBohr:2009rd,Stieberger:2009hq,Chen:2010sr}.
Gravity loop amplitudes also have simple forms and one-loop
amplitudes in ${\cal N}=8$ supergravity satisfy a 'no-triangle
property'~\cite{Bern:2005bb,BjerrumBohr:2006yw,Bern:2007xj}.
This no-triangle property was first proven
in~\cite{BjerrumBohr:2008ji,ArkaniHamed}. No-triangle type of
simplifications have been shown to hold even for pure gravity
amplitudes~\cite{Bern:2007xj,BjerrumBohr:2008ji} and also at
the level of multi-loop amplitudes. Specifically, ${\cal N}=8$
supergravity amplitudes have been shown to satisfy very
non-trivial cancellations at four-loop level~\cite{Bern:2006kd}
(see also~\cite{Green:2006gt}). Interestingly, KLT relations
can be shown to hold regardless of the external matter content
of the theory and in arbitrary dimension. KLT relations also
hold for effective field theories of
gravity~\cite{Donoghue:1994dn,Bern:1999bx,EffKLT}. We refer to
ref.~\cite{Bern:2002kj} for a recent review and for further
references on KLT and gravity amplitudes.

Although the KLT relations hold at the amplitude level in field
theory, such relations have no natural framework at the
Lagrangian level. It was however shown in a striking paper by
Bern and Grant~\cite{Bern:1999ji} (see
also~\cite{Ananth:2007zy}) that one can construct graviton
vertex interactions exclusively using the KLT relations as well
as QCD gluon amplitudes as input. In this paper it is our aim
to extend their analysis of graviton interactions to gravitinos
and demonstrate how compact and factorized gravitino Feynman
rules can be fabricated. As in the paper by Bern and Grant we
will use the KLT relations together with amplitude expressions
from Yang-Mills theory to derive results for vertex rules. As
we are considering gravitino amplitudes we need to consider
both gluon and fermion amplitudes. All results will be
demonstrated to be consistent with those derived via
traditional methods.

\section{Review of KLT relations}
Tree amplitudes in gravity theories can be computed using the
Einstein-Hilbert Lagrangian which is given by
\begin{equation}
\mathcal{L}=\frac{2}{\kappa^2}\sqrt{-g}R +{\cal L}_{\rm m}\,.
\end{equation}
We employ conventional metric $(+1,-1,-1,-1)$ and define
$\kappa=\sqrt{32\pi G}$. In the above equation $g$ denotes the
determinant of the metric and $R$ is the Ricci scalar. The term
${\cal L}_{\rm m}$ contains possible matter interactions
involving gravitons and other fields such as {\it e.g.}
gravitinos. In order to derive vertex interactions one normally
expands $g_{\mu\nu}$ around flat space, ({\it i.e.} $g_{\mu\nu}
= \eta_{\mu\nu}+\kappa h_{\mu\nu}$), make a choice of gauge,
and finally derive propagator and vertices from considering
terms of order $\sim h^2$, $\sim h^3$, $\ldots$ and various
matter couplings respectively.

In the text we will let $h$, $\widetilde{h}$, $g$ and
$\widetilde{g}$ denote gravitons, gravitinos, gluons and
gluinos. We will use the spinor helicity formalism
wherever convenient.
We define the antisymmetric tensor by $\epsilon^{0123}={+}1$
and $\gamma_5$ by $i\;\gamma_0\gamma_1\gamma_2\gamma_3$. All
momenta in amplitudes are considered to be out-going. We will define
\begin{equation}
|p^\pm\rangle=u^\pm(p)=\frac{1}{2}(1\pm\gamma_5)u(p),\ \ \  \langle
p^\pm|=\bar{u}^\pm(p)=\bar{u}(p)\frac{1}{2}(1\mp\gamma_5)\,,
\end{equation}
where $\cancel{p}u(p)=0$ and $|p^\pm\rangle=C(\langle
p^\pm|)^T$ and $C$ is the charge conjugation matrix
($C^T={-}C$, $C^\dag C=1$, $C^{-1}\gamma_\mu
C={-}\gamma^T_\mu$). Spinors are normalized so that
\begin{equation}
\langle
p^\pm|\gamma^\mu|p^\pm\rangle=\bar{u}^\pm_a(p)\gamma^{\mu\,(ab)}u^\pm_b(p)=2p^\mu\,.
\end{equation}
At times we will use shorthand notation such as
\begin{equation}
\langle AB\rangle=\langle
p_A^-|p_B^+\rangle,\ \ \ [AB]=\langle p_A^+|p_B^-\rangle\,.
\end{equation}
All spinor products are anti-symmetric so that
$\langle AB\rangle=-\langle BA\rangle$ and they are all
subject to the Schouten identities
\begin{equation}
\langle AB\rangle\langle CD\rangle+\langle AD\rangle\langle BC\rangle+\langle AC\rangle\langle DB\rangle=0\,.
\end{equation}
We have
$\langle
p_A^-|\gamma^\mu|p_B^-\rangle=\langle
p_B^+|\gamma^\mu|p_A^+\rangle\,,
$
and (via a Fierz rearrangement)
\begin{equation}
\langle p_A^{+}|\gamma_\mu|p_B^+\rangle\langle p_C^{+}|
\gamma^\mu|p_D^+\rangle = 2[AC]\langle DB\rangle
\label{Fierzing}\,.
\end{equation}
External gluinos can in the spinor helicity formalism
be represented by either
\begin{equation}
\bar{\varepsilon}_{\widetilde{g}}^\pm(p)=\langle p^\pm| \ \ \ {\rm or
}\ \ \  \varepsilon_{\widetilde{g}}^\pm(p)=|p^\mp\rangle\,.
\end{equation}
Polarization vectors for gluons~\cite{Xu:1986xb,Mangano:1990by} can be defined as
\begin{equation}
\varepsilon^{\pm\mu}_g(p,q)=\pm\frac{\langle p^{\pm}|\gamma^\mu|q^\pm\rangle}{\sqrt{2}\langle q^\mp|p^\pm\rangle}\,.
\end{equation}
Here $q$ is a reference spinor (we will at times suppress the reference spinors
in equations to avoid unnecessary cluttering of expressions). We will construct
polarization tensors of the gravitino field via
\begin{equation}
\bar{\varepsilon}^{\pm\mu}_{\widetilde{h}}(p,q)=\bar{
\varepsilon}_{\widetilde{g}}^\pm(p)\varepsilon^{\pm\mu}_g(p,q)\,,
\end{equation}
while the polarization tensors of the graviton field will be given by
\begin{equation}
\varepsilon^{\pm\mu\nu}_h(p,q)=\varepsilon_g^{\pm\mu}(p,q)\varepsilon^{\pm\nu}_g(p,q)\,.
\end{equation}
We will first consider more traditional ways of dealing with
gravitino scattering (see
e.g.~\cite{Rarita:1941mf,Freedman:1976xh,Ferrara:1976ni,Das:1976ct,Grisaru:1976vm,
Grisaru:1977px}). Our starting point will be the following
matter Lagrangian $({\cal L}_{\rm m})$
\begin{equation}\begin{split}
\mathcal{L}_{\rm m}=-\frac{1}{2}\epsilon^{\mu\nu\rho\sigma}\bar
{\Psi}_\mu\gamma_5\gamma_\nu D_\rho\Psi_\sigma+
\frac{\kappa^2\sqrt{-g}}{128}\big(2\bar{\Psi}_\mu\gamma_\nu
\Psi_\rho\bar{\Psi}^\nu\gamma^\mu\Psi^\rho
+\bar{\Psi}_\mu\gamma_\nu\Psi_\rho\bar{\Psi}^\mu\gamma^\nu
\Psi^\rho-4\bar{\Psi}_\mu\gamma^\mu\Psi_\rho\bar{\Psi}_\lambda
\gamma^\lambda\Psi^\rho\big)\,.
\end{split}\end{equation}
This Lagrangian governs the spin-${3\over2}$ gravitino field in
curved space. The covariant derivative is defined via
$D_\rho=\partial_\rho{+}\frac{1}{8}\omega_\rho^{\phantom{
\rho}mn}[\gamma_m,\gamma_n]$ and the spin connection
$\omega_\rho^{\phantom{\rho}mn}$ is the usual one.

To make a traditional computation of amplitudes with
external gravitinos, one can expand the kinetic part of the
Lagrangian (here to lowest order in the graviton field).
Discarding the antisymmetric vierbein field gives rise to the
following Lagrangian
\begin{equation}\begin{split}
\mathcal{L}_{\rm m}&=-\frac{1}{2}\epsilon^{\mu\nu\rho\sigma}
\bar{\Psi}_\mu\gamma_5\gamma_\nu\partial_\rho\Psi_\sigma
-\frac{\kappa}{4}\epsilon^{\mu\nu\rho\sigma}\bar{\Psi}_\mu\gamma_5\gamma^\alpha
h_{\alpha\nu}\partial_\rho\Psi_\sigma-\frac{\kappa}{16}\epsilon^{\mu\nu\rho\sigma}
\bar{\Psi}_\mu\gamma_5\gamma_\nu[\gamma^\alpha,\gamma^\beta]
\Psi_\sigma\partial_\beta h_{\rho\alpha}\label{lagrange}\\
&\hspace{3cm}+\frac{\kappa^2}{128}
\big(2\bar{\Psi}_\mu\gamma_\nu\Psi_\rho
\bar{\Psi}^\nu\gamma^\mu\Psi^\rho+
\bar{\Psi}_\mu\gamma_\nu\Psi_\rho\bar{\Psi}^\mu\gamma^\nu
\Psi^\rho-4\bar{\Psi}_\mu\gamma^\mu\Psi_\rho
\bar{\Psi}_\lambda\gamma^\lambda\Psi^\rho\big)\,.
\end{split}\end{equation}
From this equation it is simple to find vertex
factors~\cite{Nielsen:1978ex,Grisaru:1977zk}. {\it E.g.} if we
have two gravitinos with momentum $p'$ and $p$ (Lorentz indices
$\nu$ and $\mu$) and a graviton leg with momentum $q$ (Lorentz
indices $\alpha,\beta$) we arrive at (all factors of the
coupling constant $\frac{\kappa}{2}$ have been suppressed)
\begin{equation}\begin{split}
V_{\tilde h \tilde h h}^{\nu,\mu,\alpha\beta}(p',p,q)&=-\frac{i}{2}
\big([\gamma^\nu\eta^{\mu\alpha}-\gamma^\mu
\eta^{\nu\alpha}](p+p')^\beta\label{nytrevertex}
+[\cancel{p}-\cancel{p}^\prime]\eta^{\mu\nu}
\eta^{\alpha\beta}+\gamma^\alpha(p-p')^\beta
\eta^{\mu\nu}\\&\hspace{7cm}-2\gamma^\alpha\eta^{\mu\beta}
p^\nu+2\gamma^\alpha\eta^{\nu\beta}p^{\prime\mu}\big)\,.
\end{split}\end{equation}

\section{The Gravitino interactions from Yang-Mills theory}
\begin{figure}[!th]
\begin{center}{\includegraphics[width=17cm]{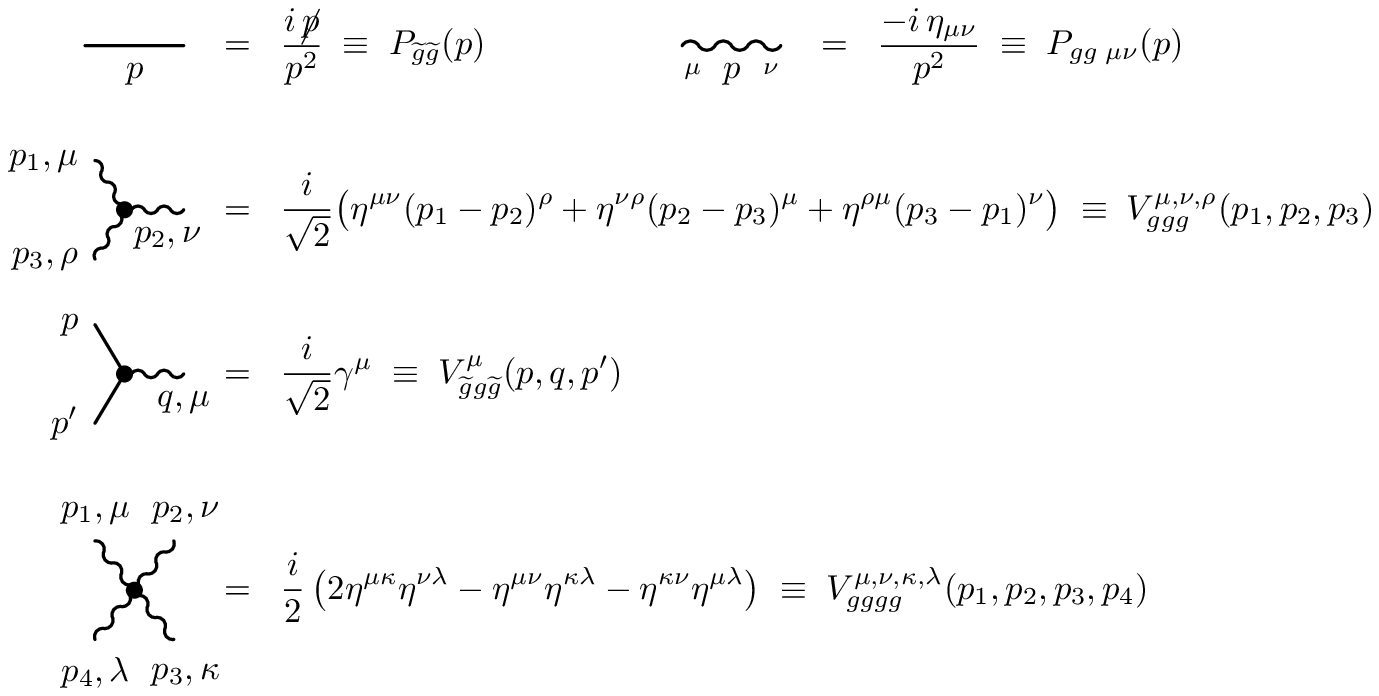}\vspace{-.8cm}
}
\end{center}
\caption{Color-ordered Feynman rules for gluons and
gluinos in Feynman gauge with all momenta out-going.\label{Feynmanregler}}
\end{figure}

We will in this section use the KLT relations to construct
(spin-${3\over2}$) gravitino Feynman rules from Yang-Mills
amplitudes involving gluons $g$ (spin-1) and
gluinos $\widetilde{g}$ (spin-$\half$).
KLT relations valid up till five points are {\it e.g.}:
\begin{equation}
\hspace{-3.35cm} \mathcal{M}_{3}{(1^{\sigma_1+\bar \sigma_1},
2^{\sigma_2+\bar \sigma_2},3^{\sigma_3+\bar \sigma_3})}\; =\; i(-1)^{N_p+1}
{\cal A}_3(1^{\sigma_1},2^{\sigma_2},3^{\sigma_3})
{\cal A}_3(1^{\bar \sigma_1},2^{\bar \sigma_2},3^{\bar \sigma_3})\,,
\end{equation}
\begin{equation}\hspace{-0.15cm}
\mathcal{M}_4(1^{\sigma_1+\bar \sigma_1},2^{\sigma_2+\bar
\sigma_2},3^{\sigma_3+\bar \sigma_3},
4^{\sigma_4+\bar \sigma_4})\;=\;i(-1)^{N_p+1}s_{12}
{\cal A}_4(1^{\sigma_1},2^{\sigma_2},3^{\sigma_3},4^{\sigma_4})
{\cal A}_4(1^{\bar\sigma_1},2^{\bar\sigma_2},4^{\bar\sigma_4},
3^{\bar\sigma_3})\,,\label{KLT}
\end{equation}
\begin{equation}\begin{split}\hspace{-0.14cm}
\mathcal{M}_5(1^{\sigma_1+\bar \sigma_1},2^{\sigma_2+\bar \sigma_2},3^{\sigma_3+
\bar \sigma_3},4^{\sigma_4+\bar \sigma_4},5^{\sigma_5+\bar \sigma_5})\;=\;
\\ & \hspace{-3cm}
i\big[s_{12}s_{34}(-1)^{N_p}{\cal A}_5(1^{\sigma_1},2^{\sigma_2},
3^{\sigma_3},4^{\sigma_4},5^{\sigma_5})
{\cal A}_5(2^{\bar{\sigma}_2},1^{\bar{\sigma}_1},4^{\bar{\sigma}_4},3^{\bar{\sigma}_3},
5^{\bar{\sigma}_5})
\\& \hspace{-3.25cm} +s_{13}s_{24}(-1)^{N_p'} {\cal A}_5(1^{\sigma_1},
3^{\sigma_3},2^{\sigma_2},4^{\sigma_4},5^{\sigma_5})
{\cal A}_5(3^{\bar{\sigma}_3},1^{\bar{\sigma}_1},4^{\bar{\sigma}_4},2^{\bar{\sigma}_2},
5^{\bar{\sigma}_5})\big]\,.
\end{split}\end{equation}
In these equations $N_p$ is a relative measure for the number of
permutations of fermions on the right hand side of the equation as compared to
the number of permutations on the left hand side, $s_{ij}=(p_i+p_j)^2$ are
generalized Mandelstam variables, $\mathcal{M}_n$ are gravity
amplitudes (with suppressed coupling constants) and ${\cal
A}_n$ are color-ordered amplitudes defined in terms of the full
amplitude as in~\cite{Mangano:1990by} (see also~\cite{Dixon:1996wi}). The propagators and vertex
factors for color-ordered gluon/gluino diagrams are shown in
fig.~\ref{Feynmanregler} in Feynman gauge. (Diagrams
can have additional minus signs due to the
fermionic nature of the gluinos. For instance, in the amplitude
of four gluinos an extra minus sign occurs when a fermionic
line connects the first and the last external leg.)

We will use the following Lagrangian for Yang-Mills theory,
\begin{equation}\begin{split}
\mathcal{L}&=\frac{i}{2}\mathrm{Tr}(\bar{\Psi}\cancel{D}\Psi)
-\frac{1}{4}\mathrm{Tr}(F_{\mu\nu}F^{\mu\nu})\,,
\end{split}\end{equation}
where
$D_\mu\Psi=\partial_\mu\Psi{+}\frac{ig_{Y\!M}}{\sqrt{2}}[A_\mu,\Psi]$
is the covariant derivative, $F_{\mu\nu}=\partial_\mu
A_\nu{-}\partial_\nu
A_\mu{+}\frac{ig_{Y\!M}}{\sqrt{2}}[A_\mu,A_\nu]$ is the field
tensor and $g_{Y\!M}$ is the coupling constant which will be suppressed
throughout this paper. $A_\mu$ denotes
here the gluon field and $\Psi$ the gluino field. The
generators of the group are normalized by
$\mathrm{Tr}(T^aT^b)=\delta^{ab}$ and $[T^a,T^b]=if^{abc}T^c$.
\subsection{Analysis of the four point gravitino amplitude}
\begin{figure}[h]
\begin{center}\includegraphics[width=10cm]{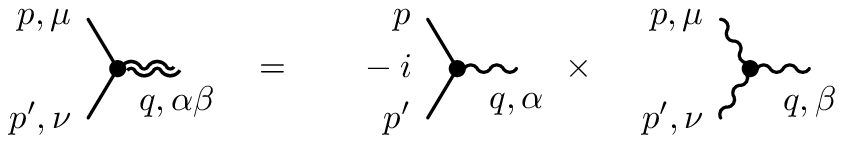}\vspace{-0.8cm}
\end{center}
\caption{Factorization of the gravitino three
vertex.\label{3ver}}
\end{figure}

Inspired by the KLT relations and the analysis by Bern and
Grant~\cite{Bern:1999ji}, we will write a factorized three
vertex for two gravitino one graviton scattering as follows
\begin{equation}\begin{split}
V^{\mu,\alpha\beta,\nu}_{\widetilde{h}h\widetilde{h}}(p,q,p')&\;=\; -i\left[\frac{i}{\sqrt{2}}
\gamma^\alpha\right]\times\bigg[\frac{i}{\sqrt{2}}\big((p-q)^\nu\eta^{\beta\mu}+
(q-p')^\mu\eta^{\beta\nu}+(p'-p)^\beta\eta^{\mu\nu}\big)\bigg]\label{trevertex}\\
&\;=\; -i\,V^{\alpha}_{\widetilde{g}g\widetilde{g}}(p,q,p')\;V^{\mu,\beta,\nu}_{ggg}(p,q,p')\,.
\end{split}\end{equation}
The vertex is also shown diagrammatically below (see fig.~\ref{3ver}).

\begin{figure}[!th]
\begin{center}\includegraphics[width=11cm]{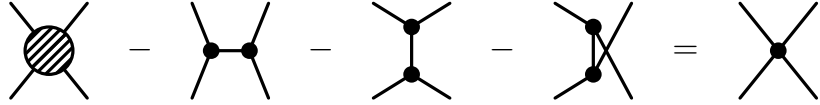}\vspace{-0.7cm}\end{center}
\caption{The scheme for obtaining the four vertex
through the KLT relations.\label{figur4}}
\end{figure}

\noindent For the propagator of gravitons we can choose the
same as in~\cite{Bern:1999ji}.
\begin{equation}
P_{hh\;\mu\alpha,\nu\beta}(p)=\frac{i\eta_{\mu\nu}\eta_{\alpha\beta}}{p^2+i\epsilon}\,.\label{propBG}
\end{equation}
Armed with a candidate for the gravitino three point vertex and
a graviton propagator we can proceed to pursue what
the four point contact interaction is. To find this we start with the four point
amplitude generated by the KLT relation. From this amplitude we subtract
all contributions involving the three vertex. What we are left with will
form the four point contact vertex contribution. The
procedure is illustrated below (see fig.~\ref{figur4}).

We begin by writing out the amplitude
$\mathcal{M}_4(1^-_{\widetilde{h}},2^-_{\widetilde{h}},
3^+_{\widetilde{h}},4^+_{\widetilde{h}})$. Using KLT we have
\begin{equation}\begin{split}
\mathcal{M}_4(1^-_{\widetilde{h}},2^-_{\widetilde{h}},
3^+_{\widetilde{h}},4^+_{\widetilde{h}})&=-i\,s_{13}\,
{\cal A}_4(1^-_g,3^+_g,2^-_g,4^+_g)\,{\cal A}_4(1^-_{\widetilde{g}},
3^+_{\widetilde{g}},4^+_{\widetilde{g}},2^-_{
\widetilde{g}})\label{langKLT}\,.
\end{split}\end{equation}
In the above expression we can investigate the pole structure of the gluino part
of the expression using the Feynman rules. We only have one
diagram because the gluino-gluino-gluon vertex is helicity
conserving. Thus we have (we have for convenience suppressed the notion of the
reference spinors in the polarization tensors and
momenta in the vertex factors as well as in the propagators.)
\begin{equation}\begin{split}
s_{13}\,{\cal A}_4(1^-_{\widetilde{g}},3^+_{\widetilde{g}},4^+_{
\widetilde{g}},2^-_{\widetilde{g}})&=s_{13}\Big(\;\bar{
\varepsilon}_{\widetilde{g}}^-(p_1)\,V^\alpha_{\widetilde{g}g
\widetilde{g}}\,\varepsilon^+_{\widetilde{g}}(p_3)\;P_{gg\;\alpha\beta}
\;\bar{\varepsilon}_{\widetilde{g}}^-(p_2)\,V^\beta_{\widetilde{g}g
\widetilde{g}}\,\varepsilon^+_{\widetilde{g}}(p_4)\Big)\\
&=s_{13}\Big(
\langle1^{-}|\gamma_\alpha|3^{-}\rangle\; \frac{i}{2s_{13}}\;
\langle2^{-}|\gamma^\alpha|4^{-}\rangle\Big)\,.
\label{s13}
\end{split}\end{equation}
One sees that the prefactor of $s_{13}$ from the KLT relations
cancels completely in this contribution.

Now let us now focus our attention to the gluon part. We will let the Lorentz
index of particle $i$ be $\sigma_i$. There are three
contributions corresponding to three distinct Feynman graphs.
\begin{equation}\begin{split}
\mathcal{A}_4(1^-_g,3^+_g,2^-_g,4^+_g)&\;=\; \varepsilon_{g\sigma_1}^-(p_1)\,
\varepsilon_{g\sigma_2}^-(p_2)\,
\varepsilon_{g\sigma_3}^+(p_3)\,
\varepsilon_{g\sigma_4}^+(p_4)\, \times \\&\hspace{0.5cm}
\Big(V^{\sigma_1,\sigma_3,\mu}_{ggg}\,
\Big(\frac{-i\eta_{\mu\nu}}{s_{13}}\Big)\,V^{\nu,\sigma_2,\sigma_4}_{ggg}
+V^{\sigma_4,\sigma_1,\mu}_{ggg}\,\Big(\frac{-i\eta_{\mu\nu}}{s_{14}}\Big)\,
V^{\sigma_3,\sigma_2,\nu}_{ggg}+V^{\sigma_1,\sigma_3,\sigma_2,\sigma_4}_{gggg}\Big)
\,.
\end{split}\end{equation}
Writing out
$\mathcal{M}_4(1^-_{\widetilde{h}},2^-_{\widetilde{h}},
3^+_{\widetilde{h}},4^+_{\widetilde{h}})$ using the above
results we have
\begin{equation}\begin{split}\hspace{-3.8cm}
\mathcal{M}_4(1^-_{\widetilde{h}},2^-_{\widetilde{h}},
3^+_{\widetilde{h}},4^+_{\widetilde{h}})&=-i\Big(
\langle1^{-}|\gamma_\alpha|3^{-}\rangle\,
\langle2^{-}|\gamma^\alpha|4^{-}\rangle\, V^{\sigma_1,\sigma_3,\mu}_{ggg}
\Big(\frac{-i\eta_{\mu\nu}}{2s_{13}}\Big)
\,V^{\nu,\sigma_2,\sigma_4}_{ggg}  \\&
\!\!\hspace{0.9cm}+\langle1^{-}|\gamma_\alpha|3^{-}\rangle\,
\langle2^{-}|\gamma^\alpha|4^{-}\rangle\,
V^{\sigma_4,\sigma_1,\mu}_{ggg}\,\Big(\frac{-i\eta_{\mu\nu}}{2s_{14}}\Big)\,
V^{\sigma_3,\sigma_2,\nu}_{ggg} \\&\!\!\hspace{0.9cm}+
\langle1^{-}|\gamma_\alpha|3^{-}\rangle\,
\langle2^{-}|\gamma^\alpha|4^{-}\rangle\, \half V^{\sigma_1,\sigma_3,\sigma_2,\sigma_4}_{gggg}
\Big)\,\varepsilon_{g\sigma_1}^-(p_1)\,\varepsilon_{g\sigma_2}^-(p_2)\,
\varepsilon_{g\sigma_3}^+(p_3)\,\varepsilon_{g\sigma_4}^+(p_4)\,.\hspace{-4cm}
\end{split}\end{equation}
The first term gives exactly a factorized gravitino vertex
structure as suggested by eq.~(\ref{trevertex}) {\it i.e.}
\begin{equation}\begin{split}\hspace{-1cm}
\langle1^{-}|\gamma_\alpha|3^{-}\rangle\,
\langle2^{-}|\gamma^\alpha|4^{-}\rangle\, V^{\sigma_1,\sigma_3,\mu}_{ggg}
\Big(\frac{-i\,\eta_{\mu\nu}}{s_{13}}\Big)\,V^{\nu,\sigma_2,\sigma_4}_{ggg}&\\ &\hspace{-3.9cm}=
\Big(i\,\langle1^{-}|\gamma^\alpha|3^{-}\rangle\,
V^{\sigma_1,\sigma_3,\mu}_{ggg}\Big)\,
\Big(\frac{i\,\eta_{\mu\nu}\eta_{\alpha\beta}}
{s_{13}}\Big)\,\Big(i\langle2^{-}|\gamma^\beta|4^{-}\rangle\, V^{\nu,\sigma_2,\sigma_4}_{ggg}\Big)\,,
\hspace{-0.5cm}\end{split}\end{equation}
but the second term does not appear to do that. This can however be mended
by applying the Fierz rearrangement~\eqref{Fierzing},
\begin{equation}
\langle p_A^{+}|\gamma_\mu|p_B^+\rangle\langle p_C^{+}|
\gamma^\mu|p_D^+\rangle=2[AC]\langle DB\rangle
=-\langle p_A^{+}|\gamma_\mu|p_D^+
\rangle\langle p_C^{+}|\gamma^\mu|p_B^+\rangle\,.
\end{equation}
Using this we can rearrange the gravitino amplitude so that
it becomes:
\begin{equation}\begin{split}\hspace{-2cm}
\mathcal{M}_4(1^-_{\widetilde{h}},2^-_{\widetilde{h}},
3^+_{\widetilde{h}},4^+_{\widetilde{h}})&= \varepsilon_{g\sigma_1}^-(p_1)\,
\varepsilon_{g\sigma_2}^-(p_2)\,\varepsilon_{g\sigma_3}^+(p_3)\,\varepsilon_{g
\sigma_4}^+(p_4)\,\times
\\&\hspace{0.3cm}\Big(\big ( i\, \langle1^{-}| \gamma^\alpha|3^{-}\rangle\,
V^{\sigma_1,\sigma_3,\mu}_{ggg}\big )\, \frac{i\,\eta_{\alpha\beta}
\eta_{\mu\nu}}{2s_{13}}\,
\big(i\,\langle2^{-}|\gamma^\beta|4^{-}\rangle\, V^{\nu,\sigma_2,
\sigma_4}_{ggg}\big)\\&-\big(i\,
\langle1^{-}|\gamma^\alpha|4^{-} \rangle\, V^{\sigma_4,
\sigma_1,\mu}_{ggg}\big)\,\frac{i\,\eta_{\alpha\beta}\eta_{\mu\nu}}
{2s_{14}}\,\big(i\,\langle2^{-}|\gamma^\beta|3^{-}\rangle\,
V^{\sigma_3,
\sigma_2,\nu}_{ggg}\big)\\
&+\frac{i}{4}\left(\langle1^{-}|\gamma_\mu|4^{-}\rangle\,
\langle2^{-}|\gamma^\mu|3^{-}\rangle\,-\langle1^{-}|\gamma_\mu|3^{-}
\rangle\,\langle2^{-}|\gamma^\mu|4^{-}\rangle\right)\,V^{\sigma_1,
\sigma_3,\sigma_2,\sigma_4}_{gggg}\Big)\,.\hspace{-5cm}
\end{split}\end{equation}
By subtracting from the amplitude the parts with the gravitino three point
vertex~\eqref{trevertex} and the propagator~\eqref{propBG}
according to our substraction scheme (see fig.~\ref{figur4})
we have now recovered the structure of the gravitino contact
term. It should be kept in mind that there is no $s_{12}$ channel diagram because
the gravitino-gravitino-graviton vertex is helicity conserving.

\begin{figure}[h]
\begin{center}\includegraphics[width=14.5cm]{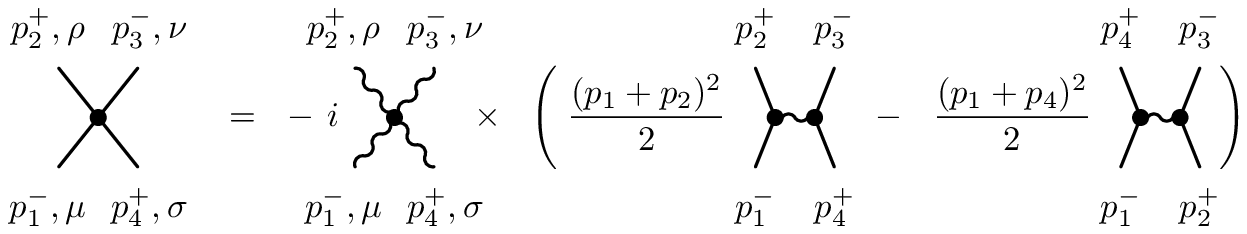}\vspace{-0.7cm}\end{center}
\caption{The factorization of the gravitino contact term.\label{kontakt}}
\end{figure}

The contact term is shown diagrammatically in
fig.~\ref{kontakt}. In the figure the helicities are specified.
We can rewrite the contact term without reference to the helicities
as follows (using eq.~\ref{Fierzing})
\begin{equation}\begin{split}
\left(\langle1^{-}|\gamma_\mu|4^{-}\rangle\,
\langle2^{-}|\gamma^\mu|3^{-}\rangle-\langle1^{-}|\gamma_\mu|3^{-}
\rangle\,\langle2^{-}|\gamma^\mu|4^{-}\rangle\right)\,V^{\sigma_1,
\sigma_3,\sigma_2,\sigma_4}_{gggg}&=\\&
\hspace{-7.48cm}\langle1^{-}|\gamma_\mu|4^{-}\rangle\,
\langle2^{-}|\gamma^\mu|3^{-}\rangle\,\frac{i}{2}(2
\eta^{\sigma_1\sigma_2}\eta^{\sigma_3\sigma_4}-
\eta^{\sigma_1\sigma_3}\eta^{\sigma_2\sigma_4}-
\eta^{\sigma_1\sigma_4}\eta^{\sigma_3\sigma_2})
\\&\hspace{-8cm}-\langle1^{-}|\gamma_\mu|3^{-}
\rangle\,\langle2^{-}|\gamma^\mu|4^{-}\rangle\,\frac{i}{2}
(2\eta^{\sigma_1\sigma_2}\eta^{\sigma_3\sigma_4}-\eta^{
\sigma_1\sigma_3}\eta^{\sigma_2\sigma_4}-\eta^{\sigma_1
\sigma_4}\eta^{\sigma_3\sigma_2})\\ &\hspace{-8cm}=\langle1^{-}|
\gamma_\mu|4^{-}\rangle\,
\langle2^{-}|\gamma^\mu|3^{-}\rangle\,\frac{i}{2}(2\eta^{
\sigma_1\sigma_2}\eta^{\sigma_3\sigma_4}-2\eta^{\sigma_1
\sigma_3}\eta^{\sigma_2\sigma_4})\\&\hspace{-8cm}-\langle1^{-}|\gamma_\mu|3^{-}
\rangle\,\langle2^{-}|\gamma^\mu|4^{-}\rangle\,\frac{i}{2}(2
\eta^{\sigma_1\sigma_2}\eta^{\sigma_3\sigma_4}-2\eta^{\sigma_1\sigma_4}\eta^{\sigma_3\sigma_2})\,.
\end{split}\end{equation}
Thus the contact term can be written (again we have here
suppressed factors of $\kappa/2$)
\begin{equation}\begin{split}
V^{\mu,\nu,\rho,\sigma}_{abcd}&=
\frac{i}{4}\Big(\!\gamma^\chi_{ab}\gamma_{\chi\,cd}
(\eta^{\mu\rho}\eta^{\nu\sigma}\!-\eta^{\mu\sigma}\eta^{\nu\rho})+\gamma^\chi_{ac}\gamma_{\chi\,bd}(\eta^{\mu\sigma}
\eta^{\nu\rho}\!-\eta^{\mu\nu}\eta^{\rho\sigma})
 +\gamma^\chi_{ad}\gamma_{\chi\,bc}(\eta^{\mu\nu}
\eta^{\rho\sigma}\!-\eta^{\mu\rho}\eta^{\nu\sigma})\!\Big)\,.
\end{split}\end{equation}
To leading order in $\bar\Psi$, $\Psi$ and $h$ one can thus instead
of~\eqref{lagrange} use the following interaction Lagrangian:
\begin{equation}\begin{split}
\mathcal{L}&=\frac{i\kappa}{4}\Big(\bar{\Psi}_\mu
\gamma^\nu(\partial^\rho\Psi^\mu)h_{\rho\nu}-\bar{\Psi}_\mu
\gamma^\nu(\partial^\mu\Psi^\rho)h_{\rho\nu}-\bar{\Psi}^\rho
\gamma^\nu\Psi^\mu(\partial_\mu h_{\rho\nu})\Big)+
\frac{\kappa^2}{64}\bar{\Psi}_\mu\gamma_\nu\Psi_\rho
\bar{\Psi}^\mu\gamma^\nu\Psi^\rho\,.
\end{split}\end{equation}

\subsection{Analysis of two gravitino two graviton
amplitude}
\noindent Let us now turn to the amplitude of two
gravitinos and two gravitons
\begin{equation}\begin{split}
\mathcal{M}_4(1^-_{\widetilde{h}},2^-_{h},
3^+_{\widetilde{h}},4^+_{h})&=-i\,s_{14}\,{\cal A}_4(1^-_{\widetilde{g}},4^+_g,2^-_g,3^+_{\widetilde{g}})
\,{\cal A}_4(1^-_g,2^-_g,3^+_g,4^+_g)\,\label{KLThhtilde}.
\end{split}\end{equation}
There are four diagrams to be considered. They are shown below
in fig.~\ref{firdia}.

\begin{figure}[h]
\begin{center}\includegraphics[width=16cm]{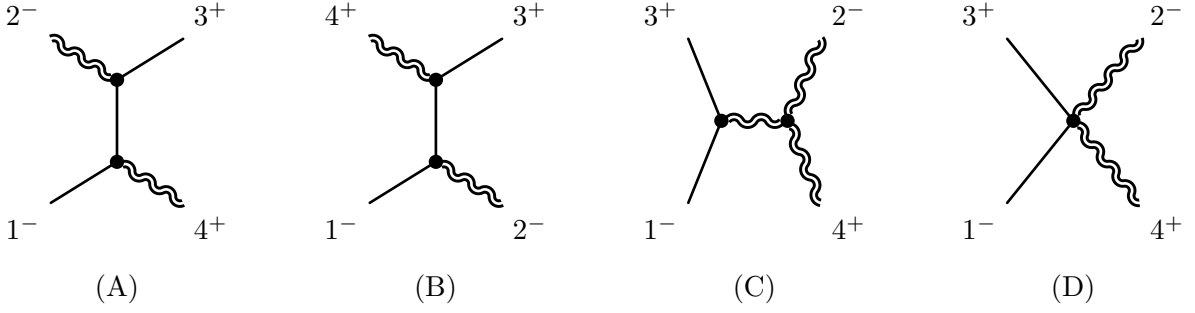}\vspace{-0.95cm}
\end{center}
\caption{The four diagrams we need to compute the amplitude
$\mathcal{M}_4(1^-_{\widetilde{h}},2^-_{h},
3^+_{\widetilde{h}},4^+_{h})$.\label{firdia}}
\end{figure}
\begin{figure}[h]
\begin{center}\includegraphics[width=8cm]{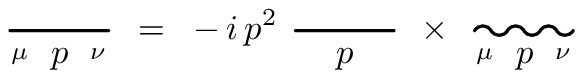}\vspace{-0.9cm}
\end{center}
\caption{The gravitino propagator.\label{prophtilde}}
\end{figure}

To calculate the diagrams (A) and (B), we need in addition to
the three vertex with one graviton leg (see~\eqref{trevertex})
a propagator for the gravitino. Inspired by~\eqref{propBG} and
the propagator used in~\cite{Das:1976ct} we choose
\begin{equation}\begin{split}
P_{\widetilde{h}\widetilde{h}\;\mu\nu}(p)&=\frac{-i\,\cancel{p}
\eta_{\mu\nu}}{p^2+i\epsilon}\,.
\end{split}\end{equation}
This propagator is shown diagrammatically in
fig.~\ref{prophtilde}.

In order to compute diagram (C) we need a graviton three vertex
in addition to the already discussed Feynman rules. Here we choose
(we have here suppressed factors of $\kappa/2$)
\begin{equation}\begin{split}
V^{\mu\nu,\alpha\beta,\rho\sigma}_{hhh}(p,q,p')&= -i \;
V^{\mu,\alpha,\rho}_{ggg}(p,q,p')\,V^{\nu,\beta,\sigma}_{ggg}(p,q,p')\\&=-i
\left(\frac{i}{\sqrt{2}}\right)^2\Big((p-q)^\rho\,
\eta^{\mu\alpha}+(q-p')^\mu\,\eta^{\alpha\rho}
+(p'-p)^\alpha\,\eta^{\rho\mu}\Big)\\&
\hspace{2.09cm}\times\Big((p-q)^\sigma\,\eta^{\beta\nu}
\label{vertexhhh}+(q-p')^\nu\,\eta^{\sigma
\beta}+(p'-p)^\beta\,\eta^{\nu\sigma}\Big)\,.
\end{split}\end{equation}
which is shown diagrammatically in fig.~\ref{3verh}:

\begin{figure}[h]
\begin{center}
\includegraphics[width=11cm]{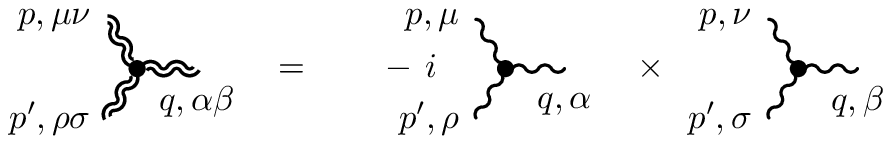}\vspace{-0.5cm}
\end{center}
\caption{The graviton three vertex.\label{3verh}}
\end{figure}

\begin{figure}[h]\hspace{-0.2cm}
\includegraphics[width=17cm]{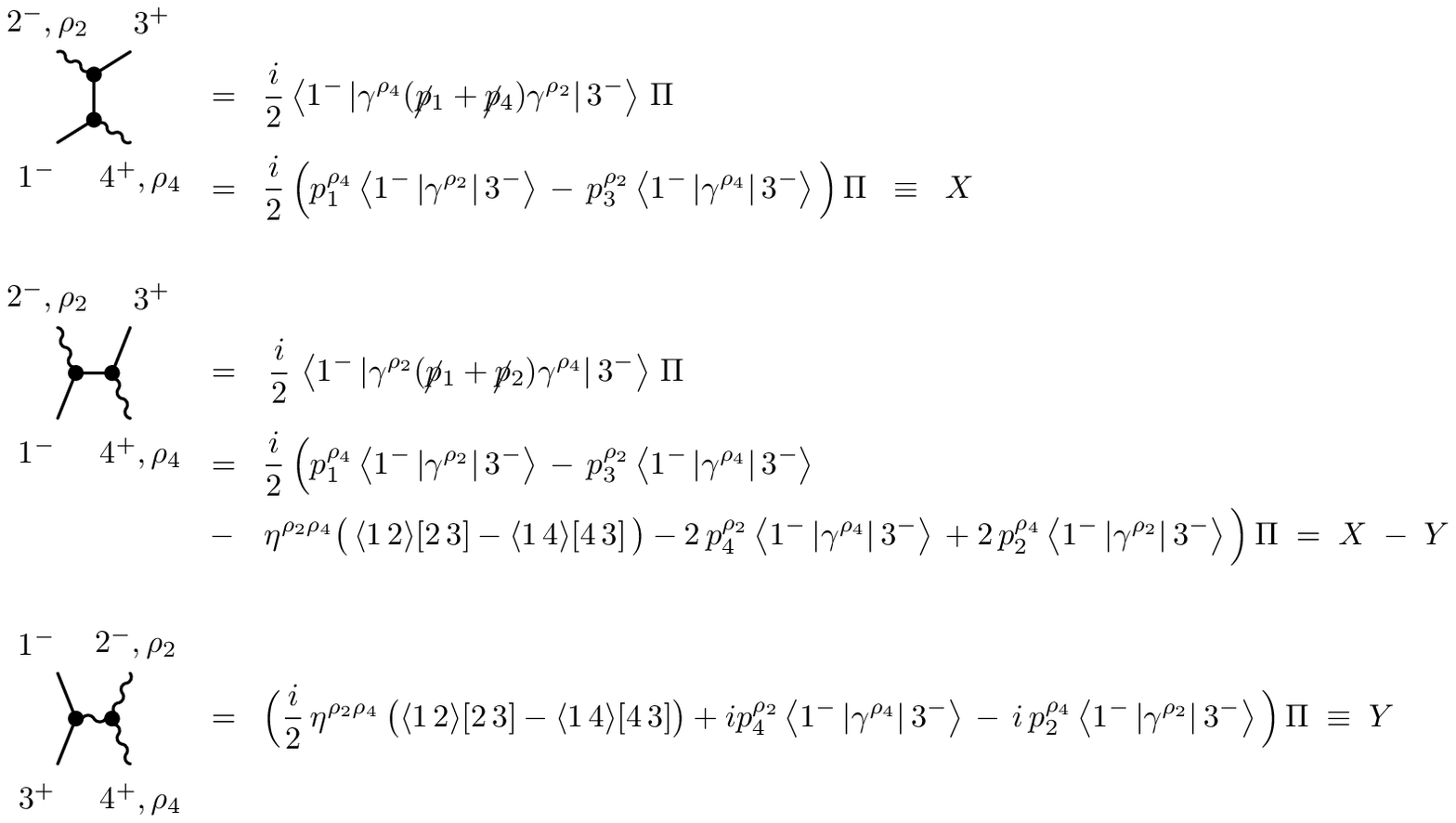}\vspace{-0.7cm}
\caption{The three different gluon/gluino diagrams with cancelled poles.\label{ggtildedia}}
\end{figure}

\begin{figure}[h]
\begin{center}
\includegraphics[width=16cm]{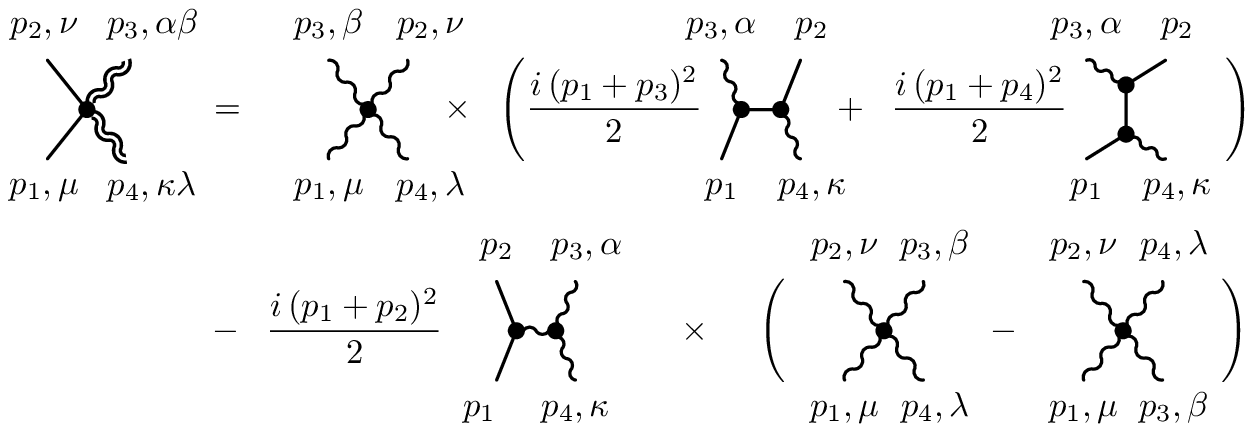}\vspace{-0.7cm}\end{center}
\caption{The factorization of the gravitino-graviton
contact term.\label{kontakthhtilde}}
\end{figure}

It is convenient to split the diagrams $(A)$, $(B)$, $(C)$ and $(D)$
into a purely gluonic diagram and a mixed gluon/gluino diagram
and let the poles be associated with the purely gluonic diagram
(we will define
$\tilde\Pi\equiv\varepsilon_{g\,\sigma_1}^-(p_1)\,
\varepsilon_{g\,\sigma_2}^-(p_2)\,\varepsilon_{g\,\sigma_3}^+(p_3)\,
\varepsilon_{g\,\sigma_4}^+(p_4)$ and $\Pi\equiv\varepsilon_{g\,\rho_2}^-(p_2)\,
\varepsilon_{g\,\rho_4}^+(p_4)$ and we have suppressed
the notion of the momenta in the vertex factors as well as in the
propagators for convenience):
\begin{equation}\begin{split}
(A)&=\tilde\Pi\,\Pi\,\langle 1^{-}|V_{\widetilde{h}\widetilde{h}h}^{
\sigma_1,\kappa,\rho_4\sigma_4}\,P_{\widetilde{h}\widetilde{h}
\;\kappa\lambda}\,V_{\widetilde{h}h\widetilde{h}}^{\lambda,
\rho_2\sigma_2,\sigma_3}|3^{-}\rangle\\
&=i\;\tilde\Pi\,\Pi\,\langle 1^{-}|\big(V_{\widetilde{g}
\widetilde{g}g}^{\rho_4}\,V_{ggg}^{\sigma_1,\kappa,\sigma_4}\big)\,
\big(s_{14}P_{\widetilde{g}\widetilde{g}}\,P_{gg\;\kappa\lambda}
\big)\,\big(V_{\widetilde{g}g\widetilde{g}}^{\rho_2}\,V_{ggg}^{
\lambda,\sigma_2,\sigma_3}\big)|3^{-}\rangle\\
&=i\;\tilde\Pi\,\Pi\,\big(\langle 1^{-}|V_{\widetilde{g}g\widetilde{g}}^{
\rho_4}\,s_{14}\,P_{\widetilde{g}\widetilde{g}}\,V_{\widetilde{g}g
\widetilde{g}}^{\rho_2}|3^{-}\rangle\big)\, \big(V_{ggg}^{\sigma_1,
\kappa,\sigma_4}\,P_{gg;\kappa\lambda}\,V_{ggg}^{\lambda,\sigma_2,\sigma_3}\big)\,.
\end{split}\end{equation}
\begin{equation}\hspace{-1.05cm}\begin{split}
(B)&=\tilde\Pi\,\Pi\,\langle 1^{-}|V_{\widetilde{h}h\widetilde{h}}^{
\sigma_1,\rho_2\sigma_2,\kappa}\,P_{\widetilde{h}\widetilde{h}
\;\kappa\lambda}\,V_{\widetilde{h}\widetilde{h}h}^{\lambda,\sigma_3,
\rho_4\sigma_4}|3^{-}\rangle\\
&=i\;\tilde\Pi\,\Pi\,\langle 1^{-}|\big(V_{\widetilde{g}g\widetilde{g}}^{\rho_2}\,V_{ggg}^{
\sigma_1,\sigma_2,\kappa}\big)\,\big(s_{12}\,P_{\widetilde{g}\widetilde{g}}\,P_{gg
\;\kappa\lambda}\big)\,\big(V_{\widetilde{g}\widetilde{g}g}^{\rho_4}\,V_{ggg}^{
\lambda,\sigma_3,\sigma_4}\big)|3^{-}\rangle\\
&=i\;\tilde\Pi\,\Pi\,\big(\langle 1^{-}|V_{\widetilde{g}g\widetilde{g}}^{\rho_2}\,s_{12}\,P_{
\widetilde{g}\widetilde{g}}\,V_{\widetilde{g}\widetilde{g}g}^{\rho_4}|3^{-}
\rangle\big)\big( V_{ggg}^{\sigma_1,\sigma_2,\kappa}\,P_{gg\;\kappa\lambda}\,
V_{ggg}^{\lambda,\sigma_3,\sigma_4}\big)\,.
\end{split}\end{equation}
\begin{equation}\begin{split}
(C)&=\tilde\Pi\,\Pi\,\langle 1^{-}|V_{\widetilde{h}h\widetilde{h}}^{
\sigma_1,\kappa\lambda,\sigma_3}|3^{-}\rangle\, P_{hh
\kappa\lambda,\alpha\beta}\,V_{hhh}^{\alpha\beta,
\rho_2\sigma_2,\rho_4\sigma_4}\varepsilon_{\sigma_1}^-(p_1)\\
&=i\;\tilde\Pi\,\Pi\,\langle 1^{-}|\big(V_{\widetilde{g}g\widetilde{g}}^{
\kappa}\,V_{ggg}^{\sigma_1,\lambda,\sigma_3}\big)|3^{-}\rangle\,
\big(s_{13}\,P_{gg\lambda,\beta}\,P_{gg\;\kappa\alpha}\big)\,\big(V_{ggg}^{
\alpha,\rho_2,\rho_4}\,V_{ggg}^{\beta,\sigma_2,\sigma_4}\big)\\
&=i\;\tilde\Pi\,\Pi\,\big(\langle 1^{-}|V_{\widetilde{g}g\widetilde{g}}^{
\kappa}|3^{-}\rangle \,s_{13}\,P_{gg\;\kappa\alpha}\,V_{ggg}^{
\alpha,\rho_2,\rho_4}\big)\,\big(V_{ggg}^{\sigma_1,\lambda,
\sigma_3}\,P_{gg\lambda,\beta}\,V_{ggg}^{\beta,\sigma_2,\sigma_4}\big)\,.
\end{split}\end{equation}
We will focus on the gluino parts of these expressions. We get (see
fig.~\ref{ggtildedia}).\\
\\
Using this we can write $(A)$, $(B)$ and $(C)$ as
\begin{equation}\begin{split}
(A)&=iX\,\tilde\Pi\,\big(V_{ggg}^{\sigma_1,\kappa,\sigma_4}\,P_{gg
\;\kappa\lambda}\,V_{ggg}^{\lambda,\sigma_2,\sigma_3}
\big)\,,\\
(B)&=i\,(X-Y)\,\tilde\Pi\,\big(V_{ggg}^{\sigma_1,\sigma_2,\kappa}\,P_{gg\;\kappa
\lambda}\,V_{ggg}^{\lambda,\sigma_3,\sigma_4}\big)\,,\\
(C)&=i\,Y\,\tilde\Pi\,\big(V_{ggg}^{\sigma_1,\lambda,\sigma_3}\,P_{gg
\;\lambda\beta}\,V_{ggg}^{\beta,\sigma_2,\sigma_4}\big)\,.
\end{split}\end{equation}
Let us now sum the contributions from the three diagrams:
\begin{equation}\begin{split}
(A)+(B)+(C)&=i\,X\,\tilde\Pi\,\big(V_{ggg}^{\sigma_1,\kappa,\sigma_4}\,
P_{gg\;\kappa\lambda}\,V_{ggg}^{\lambda,\sigma_2,\sigma_3}+
V_{ggg}^{\sigma_1,\sigma_2,\kappa}\,P_{gg;\kappa\lambda}\,
V_{ggg}^{\lambda,\sigma_3,\sigma_4}\big)\\
&+i\,Y\,\tilde\Pi\,\big(V_{ggg}^{\sigma_1,\lambda,\sigma_3}\,P_{gg\;\lambda
\beta}\,V_{ggg}^{\beta,\sigma_2,\sigma_4}-V_{ggg}^{\sigma_1,
\sigma_2,\kappa}\,P_{gg\;\kappa\lambda}\,V_{ggg}^{\lambda,\sigma_3,
\sigma_4}\big)\,.
\end{split}\end{equation}
By inspecting the terms in the parentheses, we notice that each
of them consists of the pole diagrams from color-ordered gluon
amplitudes. Hence if we add the gluon contact term
$V^{\sigma_1,\sigma_2,\sigma_3,\sigma_4}_{gggg}$ and
$V^{\sigma_1,\sigma_2,\sigma_4,\sigma_3}_{gggg}$ by hand we
arrive at
\begin{equation}\begin{split}
(A)+(B)+(C)&=i\,X\,\mathcal{A}(1_g,2_g,3_g,4_g)
-i\,X\,\tilde\Pi\, V^{\sigma_1,\sigma_2,\sigma_3,\sigma_4}_{gggg}
\\ &+i\,Y\,\mathcal{A}(1_g,2_g,4_g,3_g)
-i\,Y\,\tilde\Pi\, V^{\sigma_1,\sigma_2,\sigma_4,\sigma_3}_{gggg}\,.
\end{split}\end{equation}
If we rewrite $X$ and $Y$ in terms of the gluino/gluon diagrams
we have
\begin{equation}\begin{split}
(A)+(B)+(C)&=i\,s_{14}\,\big(\langle 1^{-}|V_{\widetilde{g}g\widetilde{g}}^{\rho_4}
\,P_{\widetilde{g}\widetilde{g}}\,V_{\widetilde{g}g\widetilde{g}}^{\rho_2}|3^{-}\rangle\big)\,
\Pi\,\mathcal{A}(1_g,2_g,3_g,4_g)\\&+i\,s_{13}\,\big(\langle 1^{-}|V_{\widetilde{g}g\widetilde{g}}^{\kappa}|3^{-}
\rangle\, P_{gg\;\kappa\alpha}\,V_{ggg}^{\alpha,\rho_2,\rho_4}\big)\,\Pi\,\mathcal{A}(1_g,2_g,4_g,3_g)\\
&-i\,X\,\tilde\Pi\, V^{\sigma_1,\sigma_2,\sigma_3,\sigma_4}_{gggg}-i\,Y\,\tilde\Pi\, V^{\sigma_1,\sigma_2,\sigma_4,\sigma_3}_{gggg}\,.
\end{split}\end{equation}
We can now use that
$s_{13}\,\mathcal{A}(1_g,2_g,4_g,3_g)=s_{14}\,\mathcal{A}(1_g,2_g,3_g,4_g)$
to rewrite
\begin{equation}\begin{split}\hspace{-0.25cm}
(A)+(B)+(C)&=i\,s_{14}\,\mathcal{A}(1_{\widetilde{g}},3_{\widetilde{g}},2_g,4_g)\,\mathcal{A}(1_g,2_g,3_g,4_g)
-i\,X\,\tilde\Pi\, V^{\sigma_1,\sigma_2,\sigma_3,\sigma_4}_{gggg}-i\,Y\,\tilde\Pi\, V^{\sigma_1,\sigma_2,\sigma_4,\sigma_3}_{gggg}\\
&=-i\,s_{14}\,\mathcal{A}(1_{\widetilde{g}},4_g,2_g,3_{\widetilde{g}})\,
\mathcal{A}(1_g,2_g,3_g,4_g)-i\,X\,\tilde\Pi\,
V^{\sigma_1,\sigma_2,\sigma_3,\sigma_4}_{gggg}-i\,Y\,\tilde\Pi\, V^{\sigma_1,\sigma_2,\sigma_4,\sigma_3}_{gggg}\,.
\end{split}\end{equation}
Thus we can read off the gravitino contact term as $i\,X\,\tilde\Pi\,
V^{\sigma_1,\sigma_2,\sigma_3,\sigma_4}_{gggg}+i\,Y\,\tilde\Pi\,
V^{\sigma_1,\sigma_2,\sigma_4,\sigma_3}_{gggg}$. That
correspond to the graviton/gravitino vertex rule (we have
again suppressed all factors of $\kappa/2$ for convenience).
\begin{equation}\begin{split}
V^{\mu,\nu,\alpha\beta,\kappa\lambda}_{\widetilde{h}\widetilde{h}hh}(p,p',q,
\widetilde{q})&=\frac{i}{8}\,\bigg[3\left(
\eta^{\mu\beta}\eta^{\nu\lambda}-\eta^{\mu
\lambda}\eta^{\nu\beta}\right)\Big((
\cancel{q}-\cancel{\widetilde{q}})\eta^{\alpha
\kappa}+\gamma^\alpha(p+p'-q)^\kappa+
\gamma^\kappa(\widetilde{q}-p-p')^\alpha\Big)\\&
\hspace{0.5cm}-\big(\gamma^\kappa(\cancel{\widetilde{q}}+\cancel{p})
\gamma^\alpha+\gamma^\alpha(\cancel{p}+\cancel{q})
\gamma^\kappa\big)\big(2\eta^{\mu\nu}
\eta^{\beta\lambda}-\eta^{\mu\beta}\eta^{\nu\lambda}-
\eta^{\mu\lambda}\eta^{\nu\beta}\big)\bigg]\,.
\end{split}\end{equation}
At the level of the Lagrangian this correspond to
\begin{equation}\begin{split}
\mathcal{L}&=\frac{3i\kappa^2}{64}\bar{\Psi}_\mu\gamma^\nu
\Psi_\rho(\partial^\chi h^{\rho\sigma}) h^{\mu\tau}
\big(2\eta_{\chi\tau}\eta_{\nu\sigma}-\eta_{\nu\tau}
\eta_{\chi\sigma}-\eta_{\sigma\tau}
\eta_{\nu\chi}\big)\\&+\frac{i\kappa^2}{128}
\bar{\Psi}_\alpha \gamma^\nu h_{\beta\nu}
\cancel{\partial}(h_{\gamma\mu}\gamma^\mu\Psi_\delta)
\times\big(2\eta^{\alpha\delta}
\eta^{\beta\gamma}-\eta^{\alpha\beta}\eta^{\gamma\delta}-
\eta^{\alpha\gamma}\eta^{\beta\delta}\big)\,.
\end{split}\end{equation}
The vertex is shown diagrammatically below
(see fig.~\ref{kontakthhtilde}).

\subsubsection{Analysis of five points, four gravitino one graviton amplitude}
\begin{figure}[h]
\begin{center}\includegraphics[width=15cm]{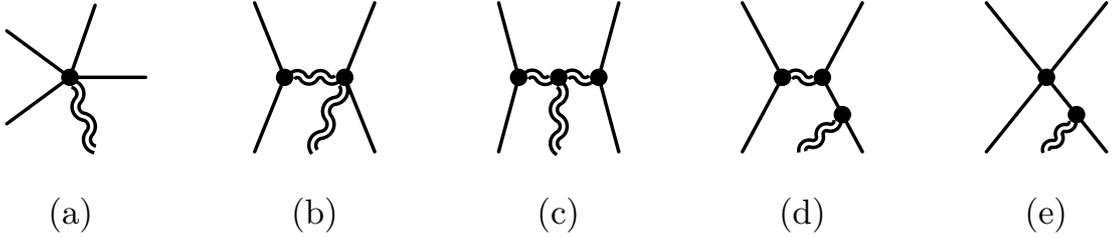}\vspace{-0.8cm}\end{center}
\caption{The five different diagram topologies for the five point function\label{femdia}}
\end{figure}

Finally, as a check of the Feynman vertex rules of the previous
section we turn to the amplitude of four gravitinos and one
graviton. There are five different Feynman graph topologies as
shown below (see fig.~\ref{femdia}).
The diagrams of topology (b), (c), (d) and (e) can be created
from the Feynman rules found in the four point case.
Contribution (a) requires the five point contact interaction.
By a careful analysis of all contributions it turns out that no
such term is necessary to get to the full amplitude. Thus the
four gravitino one graviton contact vertex factor (a) is
vanishing.

The vanishing of the five point interaction can also be understood
from the perspective of the Lagrangian given the manifest left / right mover
separated structure of such a interaction term and the odd number of Lorentz
indices which subsequently have to be contracted with each other.
However,
it is interesting that the KLT inspired formalism yields this result
with such ease. Similar arguments appear to hold for interaction
terms in the Lagrangian which have four gravitinos and any odd number
of gravitons.

\section{Conclusion}
In this paper we have extended the analysis of Bern and
Grant~\cite{Bern:1999ji} to gravitino scattering. Using
Yang-Mills theory and the KLT relations as the only input we
have derived complete Feynman rules for gravitino scattering
verified until five-point scattering with one external
graviton. The Feynman rules derived via KLT have the useful
property that they are simpler than results derived from a
conventional analysis and they are manifestly factorized into a
gluon and a gluino part. It seems clear that the organizational
principles induced by KLT rearranges the Lagrangian in useful
and simpler ways than traditional gauge choices.

Using a KLT inspired left-right separation of fields
as a way to organize and symmetrize vertex interactions
is not limited to graviton and gravitino interactions. As a
task for future research it would be very interesting to
further investigate how KLT possibly could be used to simplify
Feynman rules and computations for many other types of matter.

\begin{acknowledgments}
We would like to thank P. H. Damgaard and T. S\o ndergaard for
discussions and N. K. Nielsen for clarifying some points
concerning the article~\cite{Nielsen:1978ex}. (NEJBB) is Knud
H\o jgaard Assistant Professor at the Niels Bohr International
Academy.
\end{acknowledgments}

\end{document}